\documentclass{article}
\usepackage[utf8]{inputenc}
\usepackage{authblk}
\usepackage{setspace}
\usepackage[margin=1.25in]{geometry}
\usepackage{graphicx}
\graphicspath{ {./} }
\usepackage{subcaption}
\usepackage{amsmath}
\usepackage{lineno}

\usepackage[
style=phys,
biblabel=brackets,
abbreviate=true,
giveninits=false
]{biblatex}
\title{Recent results on heavy resonances at CMS}

\date{ Presented at the 32nd International Symposium on Lepton Photon Interactions at High Energies,\\
Madison, Wisconsin, USA, August 25--29, 2025}

\author[]{Yihui Lai$^1$~\textit{on behalf of the CMS Collaboration}}

\affil[]{$^1$ Princeton University, New Jersey, USA.}
\affil[]{Email: yihui.lai@cern.ch}

\begin{document}

\maketitle

\begin{abstract}
The Standard Model (SM) of particle physics provides a successful description of elementary particles and their interactions. However, it does not explain phenomena such as the hierarchy problem or the nature of dark matter. Many Beyond the Standard Model (BSM) theories provide answers to these open questions by adding new heavy resonances, which can be probed directly at colliders. 
This note summarizes recent resonance searches by the CMS Collaboration, focusing on final states containing top quarks and Higgs bosons. 
\end{abstract}

\section{Introduction}
Many Beyond the Standard Model (BSM) theories predict new heavy resonances that could address fundamental open questions, including the hierarchy problem and the nature of dark matter.
These resonances can decay into Standard Model (SM) particles and appear as localized excesses in invariant mass spectra above smoothly falling backgrounds, making direct searches for such signatures powerful probes of new physics.
This note presents a summary of recent CMS~\cite{CMS:2008xjf} searches for heavy resonances in top and Higgs boson final states, using the full Run~2 dataset. The results cover a broad range of spin hypotheses and mass scales, providing stringent constraints on a variety of BSM scenarios.

\section{Searches for Heavy Resonances in Top Quark Final States}

\begin{figure}[h]
    \centering
    \includegraphics[width=4.5cm]{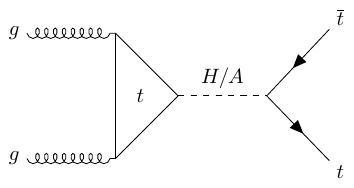}
    \includegraphics[width=3.5cm]{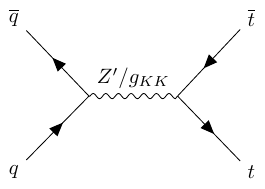}
    \includegraphics[width=4cm]{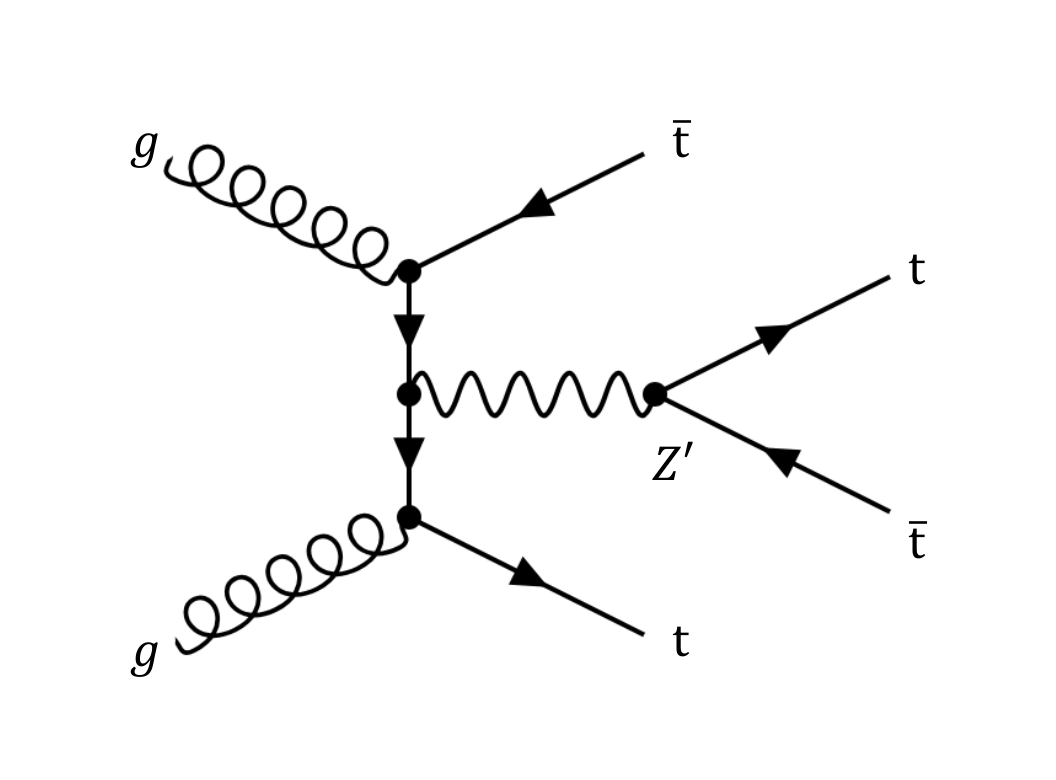}
    \caption{Representative Feynman diagrams at leading order for a scalar H or pseudoscalar A resonance production in gluon--gluon fusion (left), spin-1 particle $Z'/gKK$ production in quark--antiquark annihilation (middle), or production in association with a $t\bar{t}$ pair (right).}
    \label{fig:feynman}
\end{figure}

Neutral scalar ($H$) or pseudoscalar ($A$) bosons predicted in two-Higgs-doublet models can be produced via gluon–gluon fusion and decay into $t\bar{t}$ pairs. Spin-1 resonances such as Kaluza–Klein gluons ($gKK$), $Z'$ bosons, or dark matter mediators can be produced via quark–antiquark annihilation or in association with a $t\bar{t}$ pair. Representative Feynman diagrams for these production modes are shown in Fig.~\ref{fig:feynman}.

A search for $t\bar{t}$ resonances in the single lepton channel is presented in Ref.~\cite{CMS-PAS-B2G-22-006}.
Events are required to contain one isolated electron or muon, missing transverse momentum, and at least two jets.
Events are selected in either a resolved topology, with multiple small-radius jets, or a boosted topology, characterized by one large-radius top-tagged jet and one $b$-tagged jet. 
Jet identification is performed using the \textsc{DeepJet} and \textsc{DeepAK8} algorithms.
Signal sensitivity is enhanced using the variable $\cos\theta^{*}$, defined as the cosine of the angle between the leptonically decaying top quark in the $t\bar{t}$ rest frame and the $t\bar{t}$ system momentum in the laboratory frame.
This variable exploits differences in spin correlations between SM $t\bar{t}$ production and BSM signals. While the SM $t\bar{t}$ background peaks near $\cos\theta^{*} = \pm 1$, spin-0 resonances produce a more isotropic distribution. 

No significant excess is observed, and upper limits are set on heavy resonance production. A $Z'$ boson with widths of 1\%, 10\%, and 30\% is excluded for masses below 4.3, 5.3, and 6.7~TeV, respectively. Kaluza-Klein gluons in the Randall–Sundrum model are excluded below 4.7~TeV, and dark matter mediators below 3.2~TeV. Constraints are also set on scalar and pseudoscalar Higgs bosons with relative widths of 2.5–25\% in the 0.5–1~TeV range.

The fully hadronic $t\bar{t}$ channel~\cite{CMS-PAS-B2G-24-003} targets events where both top quarks are reconstructed as large-radius jets ($p_T>400$~GeV) identified with the \textsc{DeepAK8} tagger. The dominant multijet background is estimated using an ABCD-like method in the $(m_t, m_{t\bar{t}})$ plane. Kaluza-Klein gluons are excluded below 5~TeV, while $Z'$ dark matter mediators are excluded below 3.9~TeV. Sequential Standard Model $Z'$ bosons with 1\%, 10\%, and 30\% widths are excluded below 4.49, 5.52, and 6.85~TeV, respectively—representing the most stringent limits to date above 3~TeV.

A complementary search for $Z'\to t\bar{t}$ produced in association with a $t\bar{t}$ pair~\cite{CMS-PAS-B2G-24-009} targets four-top final states. The $Z'$ candidate is reconstructed from two large-radius jets tagged with the \textsc{ParticleNet} algorithm, while the associated tops are reconstructed in the single lepton channel. No significant deviation is observed. For  a $Z'$ width-to-mass ratio of 10\%, 20\%, and 50\%, $Z'$ masses below 564, 849, and 1125~GeV are excluded at 95\% confidence level, setting the most stringent constraints to date on a $Z'$ boson coupling exclusively to top quarks.

\begin{figure}[h]
    \centering
    \includegraphics[width=5.1cm]{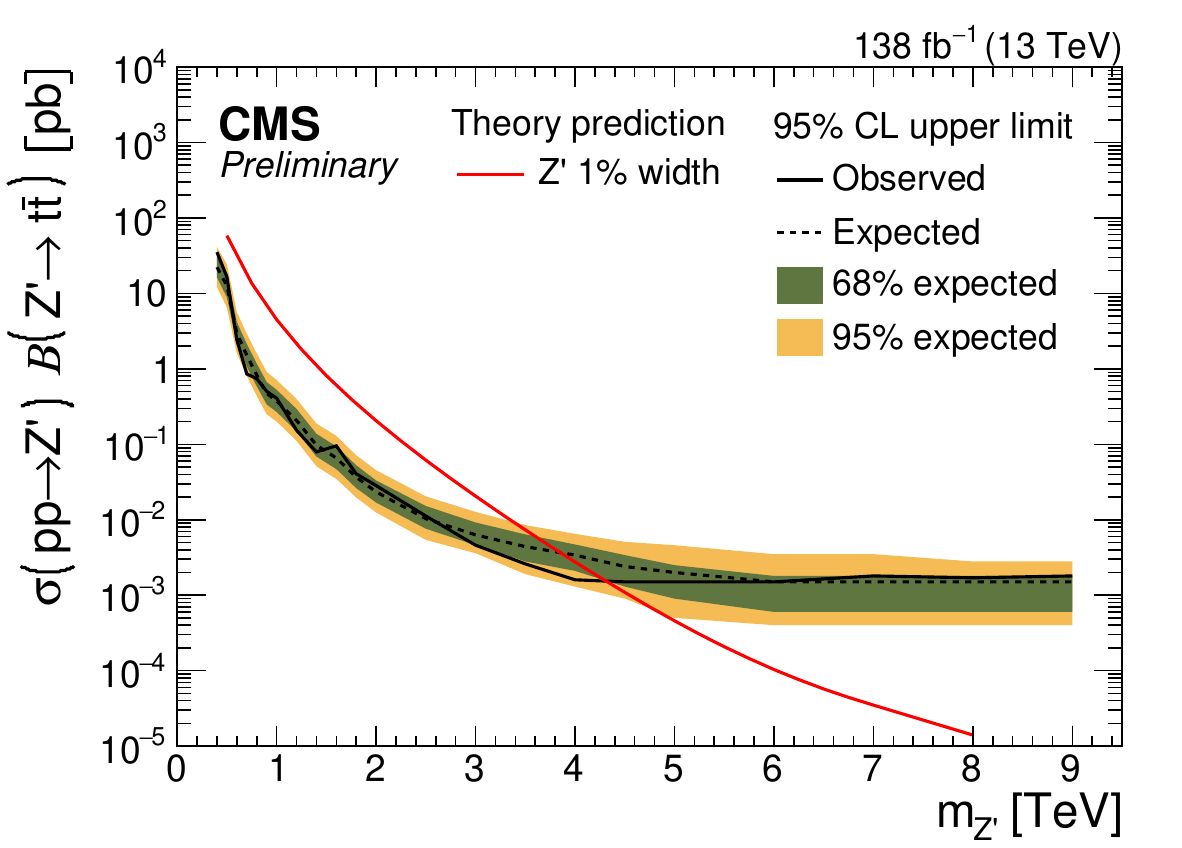}
    \includegraphics[width=4.65cm]{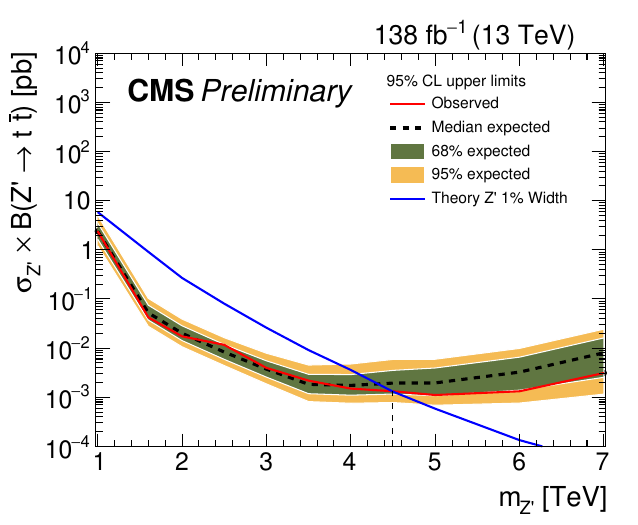}
    \includegraphics[width=5.25cm]{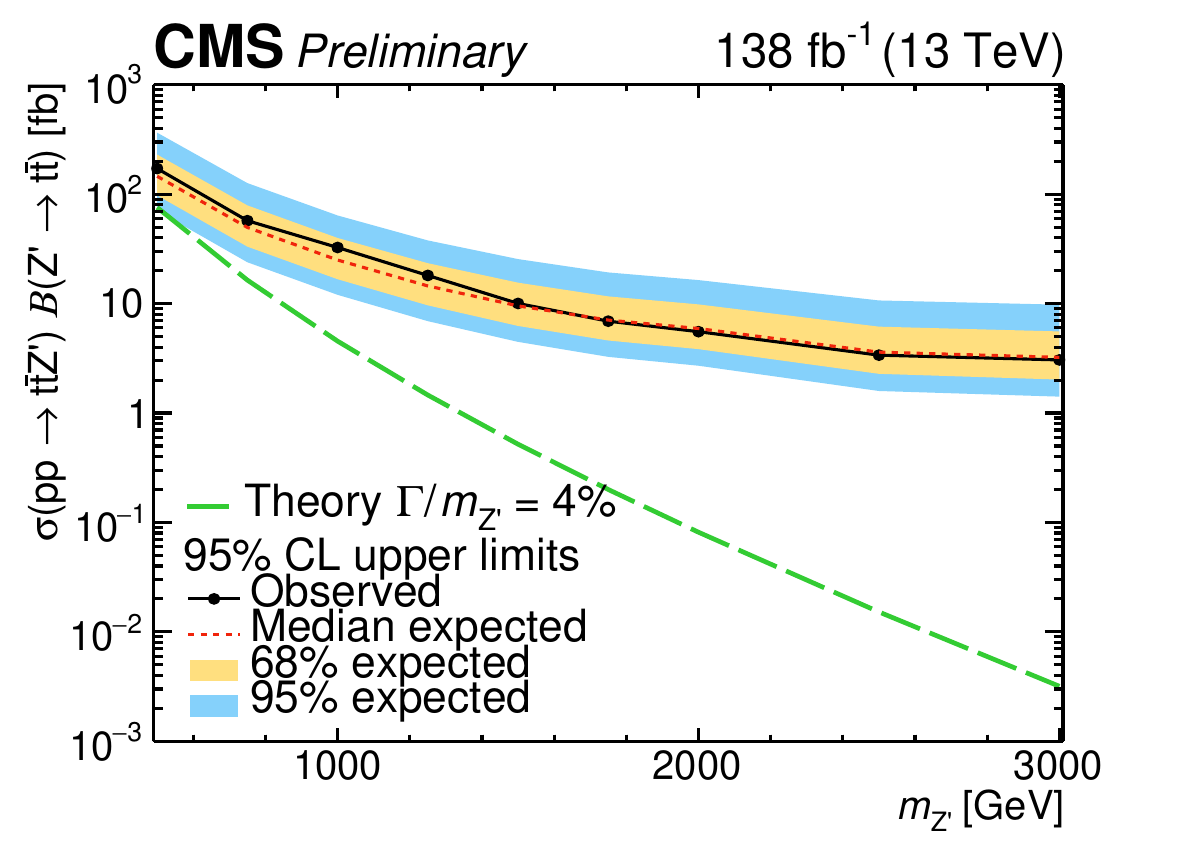}
    \caption{Observed and expected 95\% CL upper limits on the production cross section of $Z'$ bosons decaying to $t\bar{t}$. The left and middle panels show the limits for a narrow-width $Z'$ with 1\% resonance width from the single-lepton and fully hadronic analyses, respectively. The right panel shows the limits for a broader 4\% width $Z'$, obtained from the associated $t\bar{t}Z'$ production analysis~\cite{CMS-PAS-B2G-22-006,CMS-PAS-B2G-24-003,CMS-PAS-B2G-24-009}.}
    \label{fig:ttbarlimits}
\end{figure}

The observed and expected 95\% CL upper limits on the production cross section of $Z'$ bosons decaying to $t\bar{t}$ are shown in Fig.~\ref{fig:ttbarlimits}. The left and middle panels show the limits for a narrow-width $Z'$ with 1\% resonance width from the single lepton and fully hadronic analyses, respectively. The right panel shows the limits for a broader 4\% width $Z'$, obtained from the associated $t\bar{t}Z'$ production analysis.

\section{Searches for Heavy Resonances in Higgs Boson Final States}

Charged Higgs bosons are predicted in several extensions of the SM, including generalized two-Higgs-doublet models with flavor-violating couplings. A dedicated search for $H^\pm \to tb$ has been performed~\cite{CMS-PAS-B2G-24-008}.  
Events with one isolated lepton and at least two AK4 jets are categorized by jet and $b$-tag multiplicities, and a parametric deep neural network is used to enhance signal discrimination across different $H^\pm$ mass hypotheses.

\begin{figure}[h]
    \centering
    \includegraphics[width=6cm]{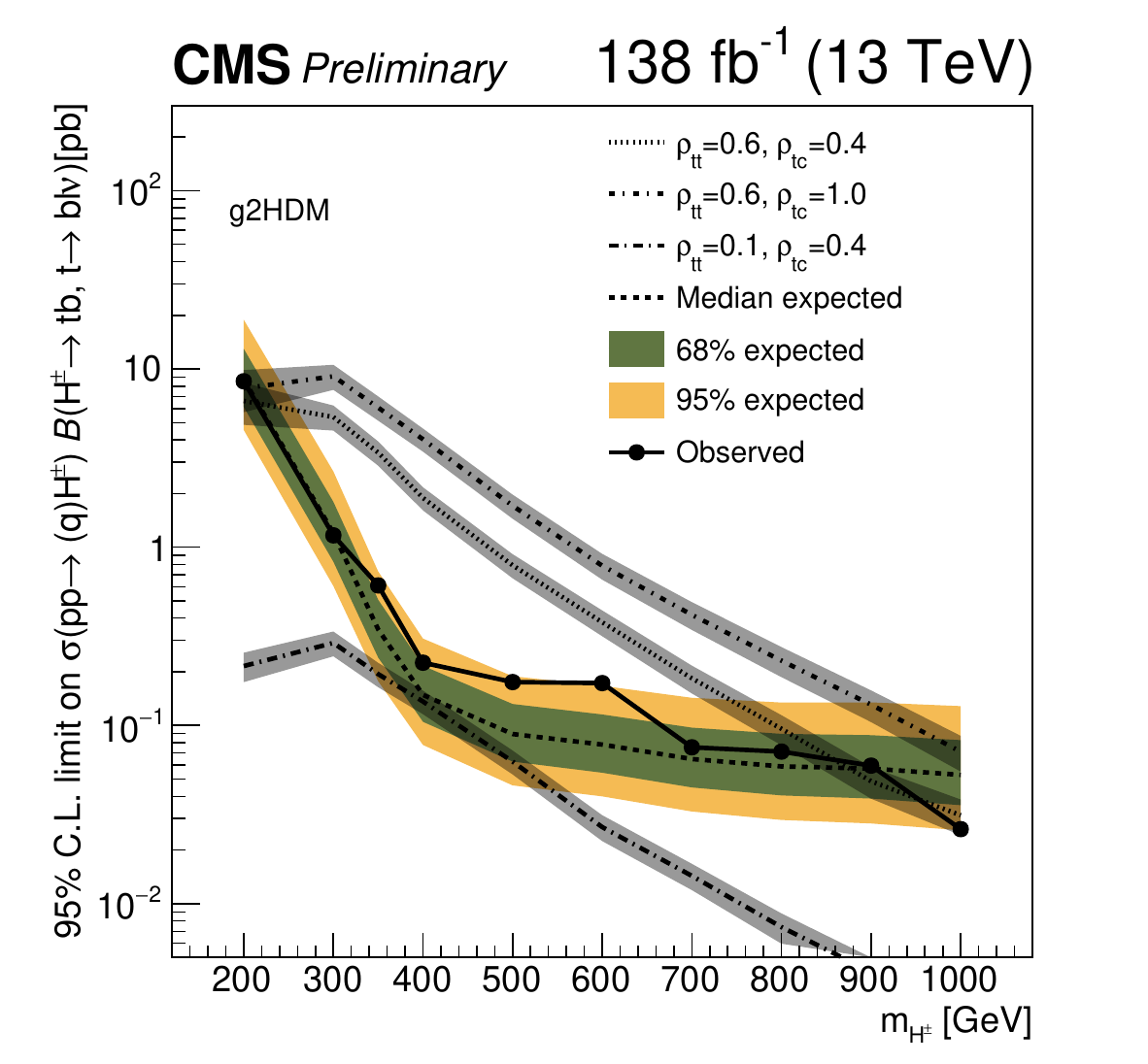}
    \includegraphics[width=6.5cm]{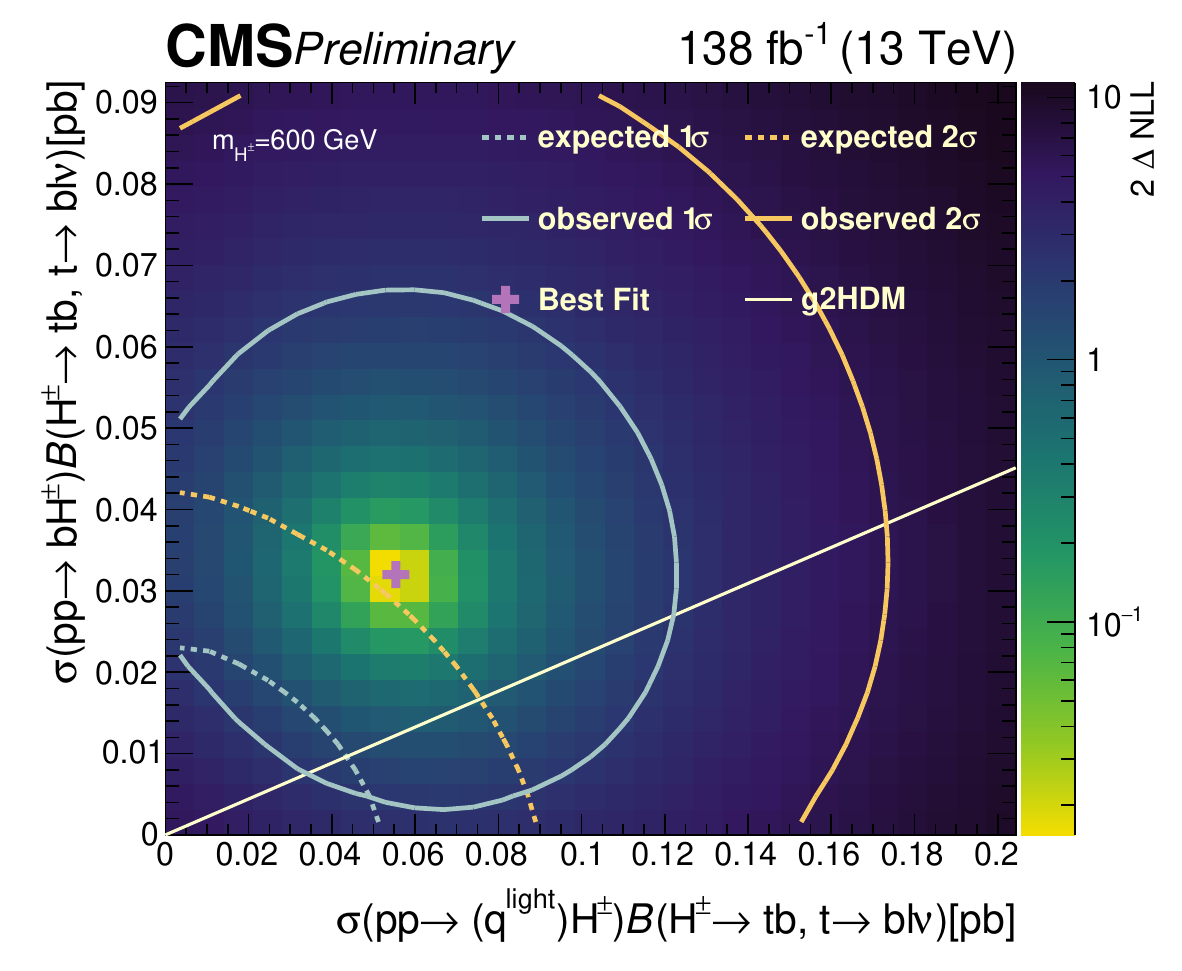}
    \caption{Observed and expected 95\% CL limits on $H^\pm \to tb$. 
    The left panel shows the mass-dependent limits. 
    The right panel shows a 2D likelihood scan for light- and $b$-quark associated production at $m_{H^\pm}=600$~GeV}
    \label{fig:hpm}
\end{figure}

The observed and expected limits on $H^\pm \to tb$ are shown in Fig.~\ref{fig:hpm} (left). A mild excess is observed near $m_{H^\pm}=600$~GeV, with a local (global) significance of $2.4\sigma$ ($0.1\sigma$). Model-independent limits are also provided by separating light- and $b$-quark associated production, allowing reinterpretation in alternative scenarios. The right panel of Fig.~\ref{fig:hpm} shows a 2D likelihood scan for light- and $b$-quark associated production at $m_{H^\pm}=600$~GeV; the best-fit value is slightly shifted from zero due to the mild local excess.

A dedicated search has been performed for heavy resonances decaying into a photon and either a Higgs or a $Z$ boson, with the boson subsequently decaying into a pair of bottom quarks~\cite{CMS-PAS-B2G-24-007}. The analysis targets resonance masses between 0.7 and 3.5~TeV and focuses on final states containing a high-$p_T$ photon and a Lorentz-boosted $b\bar b$ system reconstructed as a single large-radius jet. 

The jet mass is determined using the \textsc{ParticleNet} mass-decorrelated algorithm, and jet substructure is classified with the \textsc{Particle Transformer} tagger, efficiently identifying $H\to b\bar b$ and $Z\to b\bar b$ decays while suppressing QCD backgrounds. Events are categorized by jet mass (Z or H mass window) and tagger score (high- and medium-purity), resulting in four optimized signal regions.

Backgrounds from $\gamma$+jets and $\gamma+V$ processes are estimated directly from data via a parametric fit to the photon+jet invariant mass spectrum using a discrete profiling method. Control regions are defined by inverting the jet mass requirement, providing background-dominated samples with kinematic properties similar to the signal regions. The search is optimized for narrow resonances but also retains sensitivity to wider signals, with three benchmark fractional widths considered for the scalar resonance.

\begin{figure}[h]
    \centering
\includegraphics[width=5 cm]{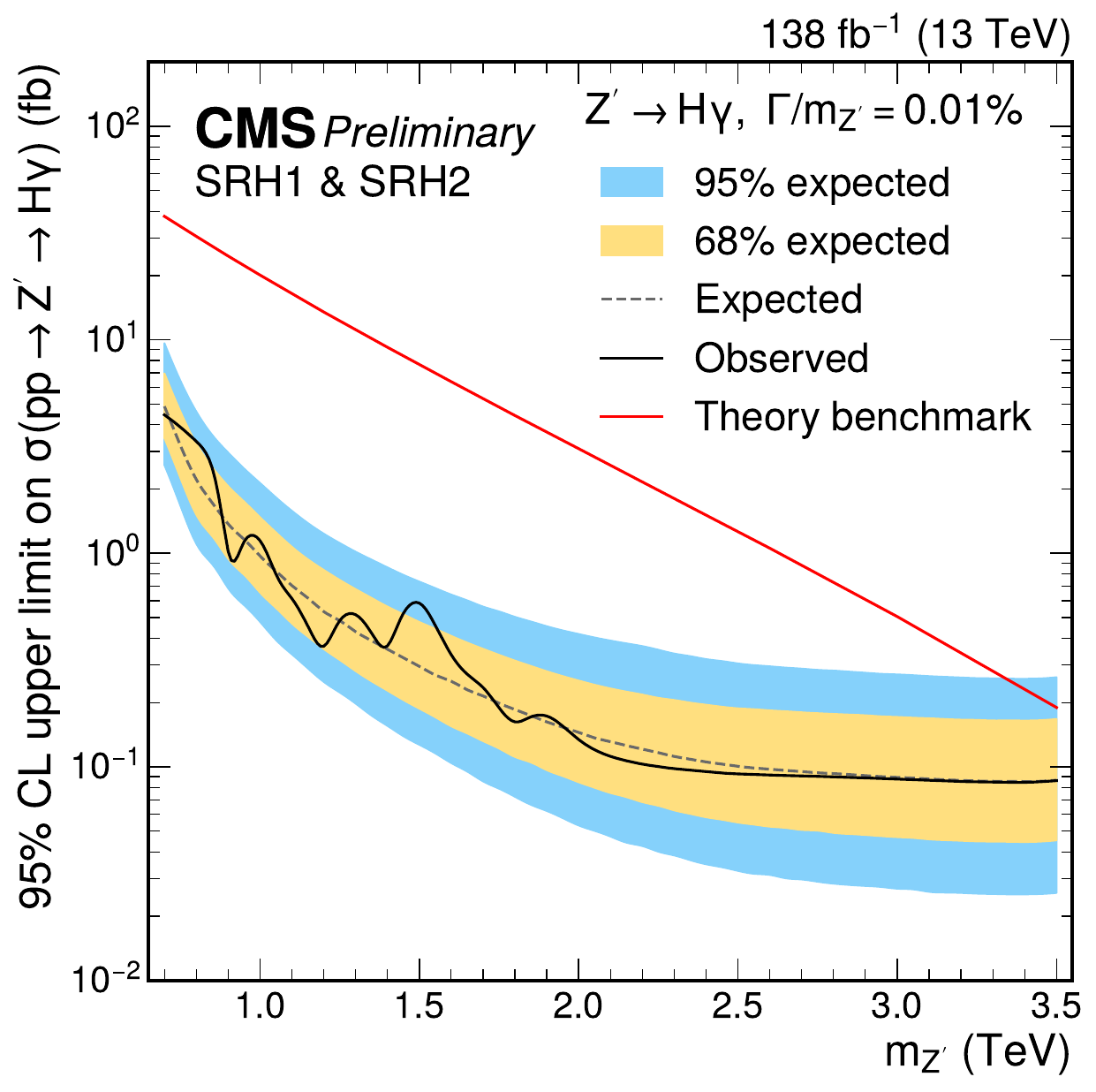}
\includegraphics[width=5 cm]{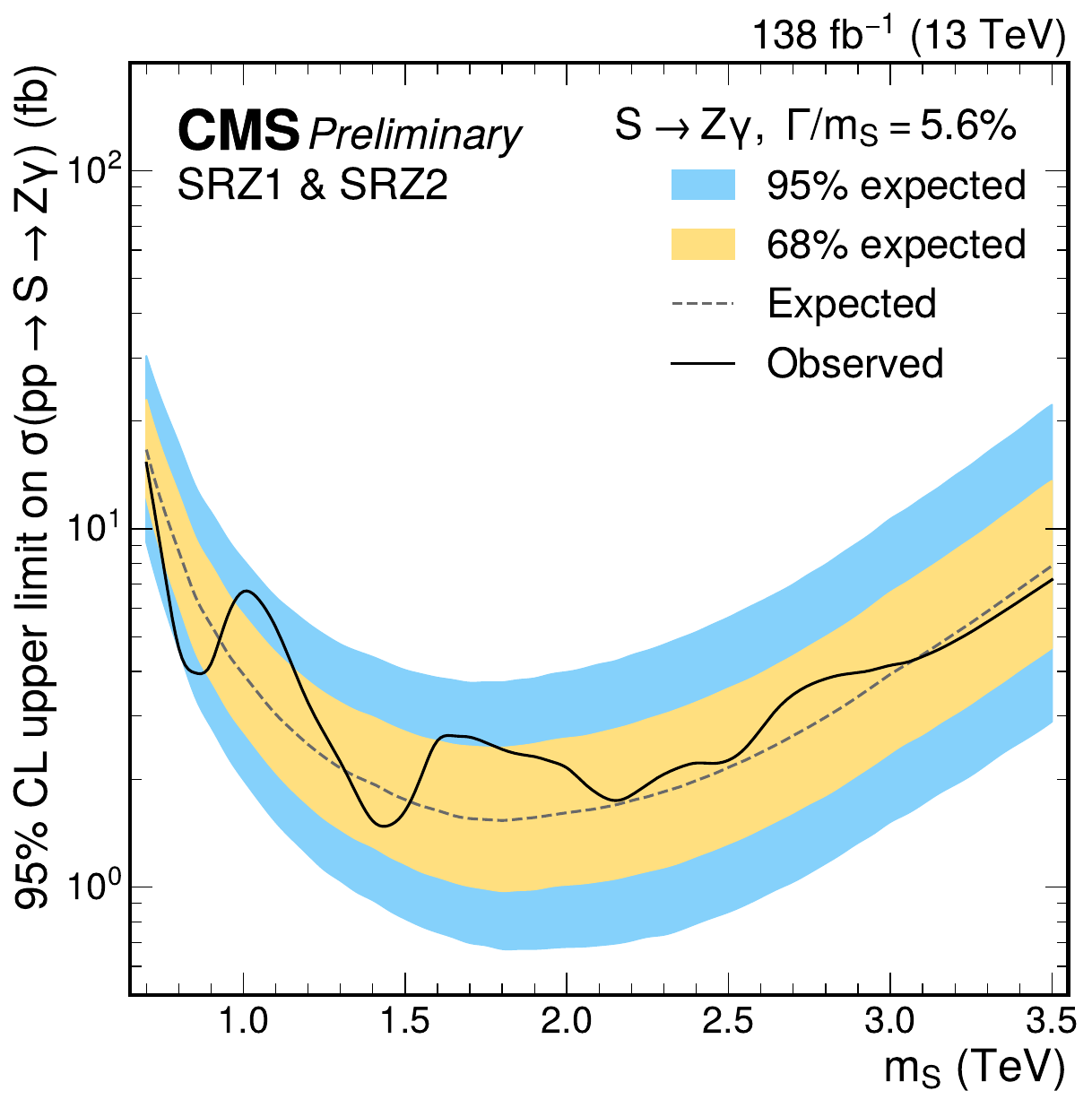}
\caption{left, expected and observed upper limits on the $Z'\to H+\gamma$, right, the upper limits on the $S\to Z+\gamma$}
\label{fig:reso}
\end{figure} 

No significant deviations from the SM background are observed. Upper limits at 95\% confidence level are set on the production cross section times branching fraction for $Z'\to H+\gamma$ and $S\to Z+\gamma$ signals across the full mass range. The observed limits improve upon previous CMS results by factors of approximately six to ten, excluding $Z'$ resonances up to 3.5 TeV and setting the most stringent constraints to date on heavy scalar $S\to Z\gamma$ decays above 1.1 TeV. These results provide a significant advance in the sensitivity of searches for exotic $V\gamma$ resonances at the LHC.

\section{Summary}
CMS has conducted comprehensive searches for heavy resonances in top quark and Higgs boson final states using the full Run~2 dataset. Neutral scalar and pseudoscalar bosons, spin-1 $Z'$ bosons, Kaluza–Klein gluons, and dark matter mediators have been probed in $t\bar{t}$ final states across single-lepton, fully hadronic, and four-top channels, employing advanced jet tagging and kinematic techniques. No significant deviations from the Standard Model are observed, leading to stringent limits: $Z'$ bosons up to 6.85~TeV, Kaluza–Klein gluons up to 5~TeV, and dark matter mediators below 3.9~TeV are excluded.
Charged Higgs bosons decaying to $tb$ are searched for with parametric deep neural network techniques, revealing a mild local excess near 600~GeV ($2.4\sigma$ local, $0.1\sigma$ global). Heavy resonances decaying to a photon and a Higgs or $Z$ boson ($H/Z\to b\bar b$) are also probed, with the most stringent limits to date excluding $Z'$ resonances up to 3.5~TeV.
These results provide the strongest constraints so far on a broad range of beyond the Standard Model heavy resonances, significantly advancing the sensitivity of LHC searches for new physics.

\printbibliography

\end{document}